# Assessment of OGC Web Processing Services for REST principles

Carlos Granell, Laura Díaz, Alain Tamayo, Joaquín Huerta

Institute of New Imaging Technologies

Universitat Jaume I de Castellón (Spain)

Avda. Vicent sos Baynat, s/n

E-12071 – Castellón (Spain)

{carlos.granell, laura.diaz, atamayo, huerta}@uji.es

## Abstract

Recent distributed computing trends advocate the use of Representational State Transfer (REST) to alleviate the inherent complexity of the Web services standards in building service-oriented web applications. In this paper we focus on the particular case of geospatial services interfaced by the OGC Web Processing Service (WPS) specification in order to assess whether WPS-based geospatial services can be viewed from the architectural principles exposed in REST. Our concluding remarks suggest that the adoption of REST principles, to specially harness the built-in mechanisms of the HTTP application protocol, may be beneficial in scenarios where *ad hoc* composition of geoprocessing services are required, common for most non-expert users of geospatial information infrastructures.

# 1. Introduction

Today, many scientists and practitioners strive to cope with heterogeneous data, services, and models with varied requirements and needs. This scenario requires the use of distributed processing capabilities and remote communications to enable collaborative and multidisciplinary research. Service-Oriented Architecture (SOA) and REpresentational State Transfer (REST) are currently the architectural styles adopted in the development of collaborative, distributed web systems and applications. Web Services based applications incarnate the SOA paradigm, while the application of REST principles yields RESTful or resource-oriented applications.

SOA is an architectural style to design applications based on a collection of best practices and patterns related to the central concept of service (Papazoglou and van der Heuvel, 2007). A service is a standards-based, loosely-coupled unit composed of a service interface and a service implementation. This conceptual duality provides a clean separation of concerns especially between public service interfaces and internal implementations. Client applications interact with SOA-based applications by specifying the desired method of a given service interface, which are mostly described using web services technology and protocols (Papazoglou, 2008) (Curbera et al., 2002). In resource-oriented applications, though, client applications interact directly with the exposed resources. The REST architectural style (Fielding, 2000) imposes a set of constraints in the communication and interaction between participants. . The application of such constraints guides the design of resource-oriented distributed systems. For instance, REST constraints are inherent to the Web architecture and materialized through the combination of HTTP, URIs and standard formats such as HTML and XML.

In the geospatial realm, Geospatial Information Infrastructures (GII) exemplify the adoption of SOA style to enable the access to distributed, heterogeneous spatial data and services through a set of common specifications and standards (Yang et al., 2010). GII may be seen as a network of multiple server nodes, which are implemented using web services technology to promote data integration and interoperability among clients and services.

Although multiple GII nodes are already deployed and running worldwide, most of them still suffer from some common issues that impede GII nodes to scale by connecting related resources and services. The reason for this difficulty might lie in the lack of connectivity between GII nodes (Díaz et al., 2011). For example, it turns out to be extremely hard for a user to find related input data sets for any given process, or to find alternative processes to a geospatial process that is down. GII is by nature an infrastructure whose aims is to provide an unique information space made up of multiples nodes interconnected, but in contrast users still face with many isolated GII nodes and geospatial services. Links and connections among the basic ingredients of GII such as geospatial data, metadata records, services and ultimately GII nodes themselves are still poor and they are in reality the exception (van Oort et al., 2010).

The paper seeks to determine whether standard geospatial services within current SOA-based GII nodes can be viewed from the REST architectural style. Taking into consideration the REST principles we analysed the particular case of geoprocessing services interfaced by the OGC[1] Web Processing Service (WPS[2]) specification (Schut, 2007), so as to assess the current state of WPS-based services with respect to the support of REST constraints. Our research objective is thus to give a technical discussion towards the application of REST principles to current geoprocessing services and geospatial services in general to promote interlinked, connected geospatial resources.

The remainder of the paper is organized as follows. Section 2 overviews the main characteristics of the WPS-based geoprocessing services. Section 3 presents the principles of REST, explains the methodology used to meet the research objective, and discusses the research hypothesis that geoprocessing services can be transformed to RESTful services to gain connectivity. Section 4 discusses some implications of the analysis undertaken compared with related work, and makes recommendations for further research on exploring RESTful geoprocessing services. Section 5 concludes the paper and points out future developments.

## 2. Geoprocessing services

Geospatial web services are central pieces in GII nodes. They allow users to access, manage, and process geospatial data in a distributed manner because these services are described in terms of standard OGC interfaces. Among them, the OGC Web Processing Service (WPS) specification (Schut, 2007) is rapidly turning in the specification of choice for exposing geospatial processing services. In this section, we describe the relevant characteristics of WPS-based services since this specification is the focus of our analysis in Section 3.

### 2.1. OGC WPS overview

OGC WPS specification provides the service interface definitions to specify a wide range of processing tasks as geospatial web services in order to distribute common GIS functionalities over the Internet (Friis-Christensen et al., 2009; Li et al., 2010). The main difference between desktop functionalities and WPS services is that the latter can be accessed remotely and assembled in varied web integration scenarios (Brunner et al., 2009; Lowe et al., 2009).

The OGC WPS provides access to calculations or models that operate on spatially-referenced data, which can be available locally, or delivered across a network using download services such as WFS (Web Feature Services[3]), WCS (Web Coverage Services[4]) and SOS (Sensor Observation Services[5]). While most OGC specifications and standards are devoted to geospatial data models and access, the OGC WPS specification is focused on processing heterogeneous geospatial data. The typical steps consist of the identification of spatially-referenced data required, execution of the process, and the management of the output process by client applications.

---

[1] Open Geospatial Consortium, http://www.opengeospatial.org

[2] The analysis conducted here is based on the last published version (1.0.0) of the WPS specification. Efforts on defining the version 2.0 are ongoing at time of writing this article, but not official yet and so subject to possible changes.

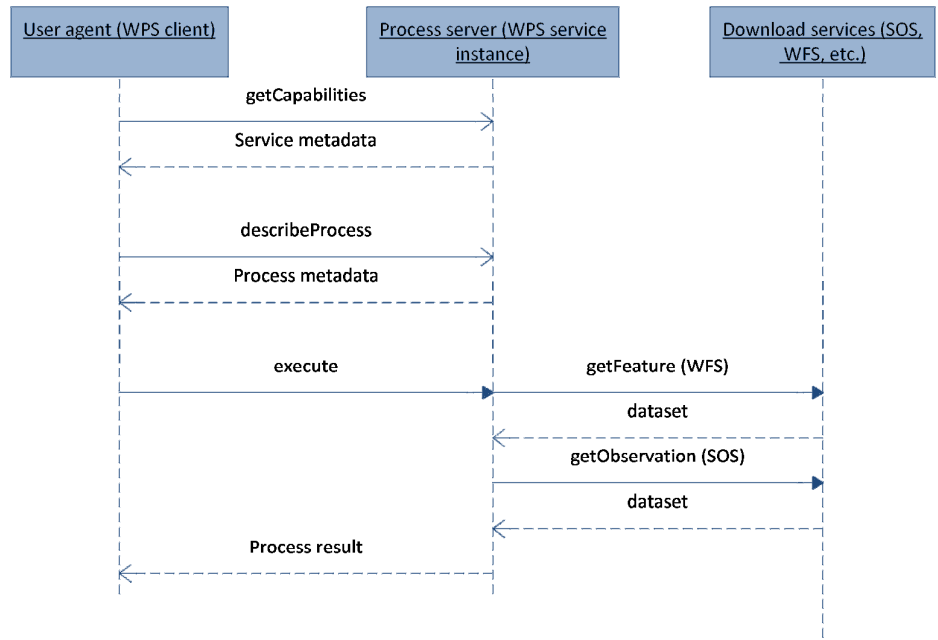

Figure 1. Request-response interactions between a WPS-compliant client and a WPS service instance

The basic operational unit is the notion of geospatial process or operation. A given WPS service instance (a concrete WPS service running) may offer one or various operations (or processes) as normal web services do. Figure 1 shows how a WPS-enabled user agent (Schaeffer and Foerster (2008) presented more elaborated WPS-aware clients) communicates with a WPS service instance, issuing three types of requests initiated by an user agent[6]: a *getCapabilities* request provokes a XML document response that contains service metadata such as server provider, contact information, and a list of available geoprocessing operations (processes) offered by the queried WPS instance; a *describeProcess* request gets as a response a XML document with detailed information for the solicited process, such as input and output parameter names and types, so that the user agent may later build the *execute* request, which eventually allows user agents to run a geospatial process.

[6] We refer hereafter to user agent as any software component or user that takes the role of client.

**2.2 Binding styles**

The term binding refers to the concrete mechanism used by user agents to access and interact with remote services. The OGC WPS specification describes two binding styles or protocols: HTTP and SOAP/WSDL. HTTP binding is widely used in current service implementations because it is the binding of choice in most OGC service specifications. HTTP requests may be issued via GET or POST methods. In the former case, request parameters are simply encoded in the same URI provided as Key-Value Pairs (KVP). This mechanism is also known as URI tunneling (Webber et al., 2010) because method names and input parameters are transferred within the URI itself. In the latter case, though, request parameters are supplied in an XML document within the HTTP body entity. This style is similar to the Plain Old XML (POX) approach where HTTP POST requests and responses are the means to transfer XML documents between clients and servers (Weber et al., 2010).

The second binding style is identical to the HTTP POST approach but with an additional SOAP envelope. In essence, SOAP binding means to package requests within a SOAP envelope through a HTTP POST method. The SOAP envelope describes a message exchange mechanism made up of a SOAP *body* element, which wraps XML documents, and a SOAP *header*. The use of SOAP does not provide additional benefits compared to the POX approach unless the SOAP *header* element is used to accommodate security certificates and encryption features to the client–server communication (Villa et al, 2008a; Villa et al., 2008b).

In practice, most WPS implementations use HTTP GET/POST communication mechanism because SOAP binding is scarcely supported and used in OGC services. Throughout this paper we will use the terms KVP binding (HTTP binding with GET KVP or URI tunneling), XML/POST binding (HTTP binding with POST XML or HTTP POX), and XML/SOAP binding (HTTP binding with POST XML plus SOAP), as they are widely used and understood by the OGC community.

# 3. Assessment of WPS services

In this section we analyze whether WPS-based services may be aligned to the set of REST constraints and elements. The list of REST constraints can vary in terminology (Fielding, 2000; Richardson and Ruby, 2007; Webber et al., 2010), but we based our discussion on the original source. According to Fielding (2000), the REST architectural style is derived from the combination and cohesion of the set of constraints listed in Section 3.1. However, the use of all of these constraints is not always necessary and suitable for the target system, depending ultimately on the concrete application requirements. Some of these constraints, uniform interface and stateless communication, are based in turn on specific REST elements which are discussed in Section 3.2. Essentially the aggregation of these constraints and elements guides developers in building distributed applications that harness the benefits of interconnected web resources.

The rest of this section evaluates each REST constraint (Section 3.1) and element (Section 3.2) as follows: first, a brief introduction to the concept is presented; then, we evaluate whether current WPS-based geoprocessing services meet or fail that REST constraint or element; and finally some design and implementation remarks towards restful geoprocessing services are suggested.

### 3.1 REST constraints

#### 3.1.1 Client-server

A service exposes a set of methods while a user agent sends requests to the service to access or run a given method. In this scenario, the client-server constraint leads to the separation of concerns, where functional capabilities are in the server side while user interface functionality is delegated to the user agent at the client side. The benefit is that implementations of services and user agents evolve independently.

From the WPS perspective, the client-server constraint is achieved since WPS-based clients and services are based on the Web service interaction paradigm (Papazoglou, 2008), which promotes a clear distinction between consumers (WPS-based clients) and providers (WPS service instances) in terms of roles and functionalities. The addition of a new process to a WPS service instance does not imply a change in the user agent in terms of new capabilities or user interface functionality, because the service interface remains intact. The WPS specification deals only with the description of service operations and the encoding of request and response messages. The standardized mechanism to access and execute every process contained in a given WPS service instance is guaranteed by the WPS specification itself.

Another question is whether the user agent is able to manage with the formats of input and output parameters imposed by a new process. However, it is worthwhile noting that the WPS specification only defines the access mechanism to processes, leaving to the service provider the definition of input and output parameters in terms of data encodings and formats. In summary, WPS-based applications take into consideration the client-server constraint.

#### 3.1.2 Stateless

Statelessness basically means that no session or application state is stored on the server side. Each user agent request to the server must contain all of the needed information so that services may understand the goal of the request without referring to any stored, shared context on the server. In REST, user agents are responsible of maintaining the application state.

OGC service communication is stateless in nature (Percivall, 2002). In general a service interaction follows the request-response pattern with no dependence on previous interactions. WPS-based services also act in this way. A user agent is able to figure out the inputs and outputs parameters for any given process from the *describeProcess* document. Then, in the subsequent *execute* interaction, user agents provide all the input parameters to perform the process. From this perspective, WPS interactions (see Figure 1) are isolated from each other.

In contrast to single executions, the combination of various processes to form a complex task or workflow normally involves previous knowledge at design time so that a designer may put together all the needed pieces. Nevertheless, the service composition design depends on specific application requirements, and it is completely independent of the WPS specification itself. That is, communications between user agents and services in the realm of WPS specification suit the stateless constraint.

### 3.1.3 Cache

The aim of the cache constraint is to improve network efficiency by eliminating some interactions between user agents and servers. When responses are cacheable, user agents can reuse them rather than issuing again equivalent requests.

The ability to selectively activate cacheable response is not considered in the WPS specification. Caching would be beneficial when clients retrieve static information. For instance, *getCapabilities* and *describeProcess* responses do not often change because they contain stable information over time. Potentially, cache-enabled user agents would perform several *execution* requests without the burden of *getCapabilities* and *describeProcess* interactions. In this scenario, user agents would have to incorporate cache components and servers would have to indicate explicitly that a response is cacheable or non-cacheable.

Fortunately, the HTTP protocol already provides cache mechanism with the use of HTTP headers such as *Cache-Control*, *Last-Modified* and *ETag* (Fielding et al., 1999). The proper combination of these headers in the client-server communication would enable the reusing of certain responses and also determine when a "fresh" response is necessary because a cacheable response has become stale or non-valid.

### 3.1.4 Uniform interface

The idea behind the uniform interface constraint is the application of the principle of service interface. Resources in the server side expose a generic interface derived from the semantics of the HTTP methods. REST raises HTTP to the level of application protocol. The purpose and meaning of the HTTP methods is meant for manipulating any resource. For instance, the GET method is for retrieving resource representations, the POST method for creating new resources, PUT for updating resources, and finally, the DELETE method to eliminate a given resource.

For each protocol binding, the WPS specification defines three operations (*getCapabilities, describeProcess, execute*) which can be encoded using either HTTP GET or HTTP POST request-response mechanism. In both cases, HTTP is the underlying transport protocol to perform these operations: an operation is either tunnelled through the *request* parameter within URI (e.g. *request=getCapabilities*) or through a XML document payload (left side Fig. 2).

Whether user agents choose KVP or XML/POST (XML/SOAP) the semantic of the operations resides in the URI or inside the XML payload, respectively. This is known RPC (Remote Procedure Call) and HTTP is merely used as a synchronous transport protocol for convenience. This breaks the uniform interface constraint since HTTP protocol is not used as an application protocol, where the standard set of HTTP methods defines the semantics of the common set of actions available to manipulate the exposed resources (right side Fig. 2).

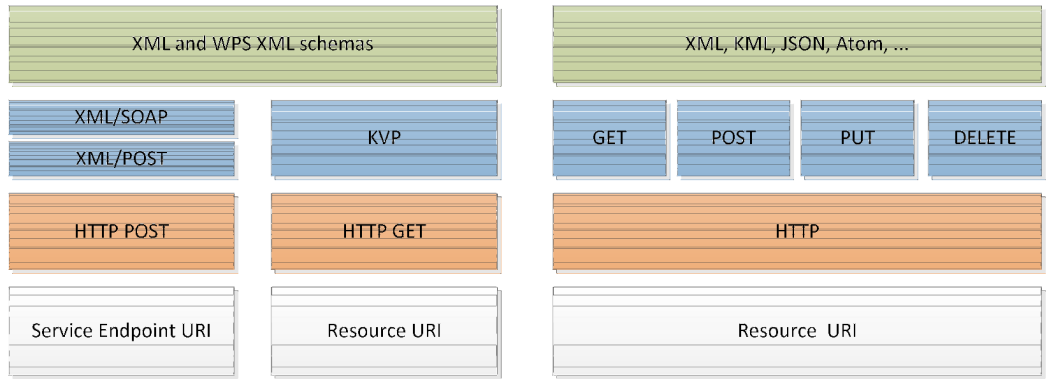

Figure 2. (Left) HTTP as transport protocol; (Right) HTTP as application protocol exposed a uniform interface

Whilst *getCapabilities* and *describeProcess* are operations to retrieve information, an *execute* operation may change the state of the server. Tunneling *execute* operations through GET violates the safety property of the GET method as defined in the HTTP protocol specification (Fielding et al., 1999). Likewise, using POST for retrieving service descriptions does not help in maintaining the principles of safety and idempotency, since POST by definition is an unsafe, i.e., it may change the state of the target resource, and it is also not idempotent because the repetition of the same request may lead to different responses. In this sense, GET and POST requests in WPS interactions are in some cases used incorrectly.

The same discussion for XML/POST may be applied when adding a SOAP envelope to wrap the XML payload (left side Fig. 2). XML/SOAP binding also treats HTTP as a transport protocol, and uses POST for any request against the same service endpoint URI. As in the case of XML/POST, the simply overuse of POST violates the uniform interface tenet.

In summary, OGC services in general follow a RPC style and uses HTTP as a transport protocol, which breaks with the concept that HTTP is neither RPC nor a transport protocol (Fielding, 2000). We will further discuss on this topic in section 3.2 as the uniform interface constraint is based on a set of REST data elements.

### 3.1.5 Layered approach

This constraint is typically regarded in distributed architectures where functionality is hierarchically decomposed on logical layers. Each layer is independent and interacts only with the immediate layer (Voisard and Schweppe, 1998). The layered approach promotes encapsulation, encourages intermediary components in separated functional layers, and minimizes the overall complexity of the system.

GIIs are multi-layered architectures (European Parliament and Council, 2007). Geospatial services play the role of mediators (Wiederhold, 1992) between client applications and data and metadata repositories. Within REST, mediators can actively transform the content of messages because these are self-descriptive (Section 3.2). From this perspective, WPS-based services may act as mediators (e.g. middleware services) by obtaining input data from remote services in order to run a process (see Figure 1). Nevertheless, this functionality is restricted to access data via HTTP requests. Advanced data manipulations such as caching, filtering or transforming the content of the messages are not supported. So, WPS services take into consideration the layered approach but it is not fully exploited. As we will describe in Section 3.2, this restriction is only partially achieved due to the lack of self-descriptive messages.

### 3.1.6 Code-on-demand
Within REST, client functionality may be extended by downloading and executing code at the client side. In this sense, responses may be uncompleted but accompanied by related executable code that altogether suits the client requirements (Erenkantz et al., 2007).

The Code-on-demand constraint is not regarded in the WPS specification, though, this dichotomy –algorithm to data or data to algorithm– has been largely discussed in the geoprocessing service literature (Müller et al., 2010; Granell et al., 2010). On demanding functions like data filtering, aggregation or fusion would be useful and valuable for user agents to manipulate service responses.

## 3.2 REST data elements
Apart from the architectural constraints, REST also relies on a set of data elements that shape the architectural constraints previously discussed. In this section, we analyse these data elements that affect some REST constraints as follows (Fielding, 2000).. Resources and identification of resources (Section 3.2.1), manipulations of resources through representation (Section 3.2.2), and hypermedia (Section 3.2.3.) shape the uniform interface constraint (Section 3.1.4); whereas self-descriptiveness (Section 3.2.4) is derived from the statelessness (Section 3.1.2) and layered approach (Section 3.1.5) constraints.

As data elements are less abstract than REST constraints, the analysis conducted in the following is accompanied with practical examples. In particular, we selected a self-developed WPS-based service[7] for topology operations that contains processes such as intersection, calculation of areas and bounding box computation for simplicity in the following explanation.

### 3.2.1 Resources and identification of resources

---

[7] http://elcano.dlsi.uji.es:8080/topologywps/WebProcessingService?request=getcapabilities&service=wps

What is considered a resource itself and how to identify it? From the REST viewpoint, the first question is straightforward since a resource can be any information that it is worth sharing with the community (Fielding, 2000; Richardson and Ruby, 2007). For instance, the Web itself functions in this way since it contains a great deal of heterogeneous resources (web page, doc, image, video, etc). Then, every single web resource is also univocally identified by its URI (Uniform Resource Identifier). The uniform interface constraint relies on the URI mechanism to identify any resource of interest so that user agents using its URI are able to access them (i.e., dereferencing).

XML/POST and XML/SOAP protocol bindings do not provide an explicit, univocal resource identification as REST encourages. A WPS service is addressable by a URI that refers to a service endpoint. Conceptually, the service acts as a proxy, receiving all incoming requests and bypassing access to the internal processes. For instance, all target topology algorithms have always the same URI that points to the WPS service endpoint, which centrally manages contained resources through service methods, such as *getCapabilities*, *describeProcess*, and *execute*. In practice, processes at the backend are hidden to user agents and thus connections can be only established to the visible proxy service exemplified by the whole WPS service.

In contrast, the KVP approach provides distinct URI for each resource. For the same WPS service, each operation has a different URI over the set of WPS processes. Whether issuing a *describeProcess* over the 'Area' process or over the 'Intersect' process both URIs are distinct.

Nevertheless, a clear limitation is that target processes according to the WPS specification use URN (Uniform Resource Name) identifiers. URN is a subset of URI yet it does not need to be linked to an existing online resource. A key difference in REST style is that the resource identifier is used for dereferencing the very resource, that is, the URI is meant to access the resource (Granell et al., 2011). For instance, the 'Area' process has the identifier 'org.n52.wps.server.algorithm.topology.Area', accessible from *getCapabilities* response document. However, user agents need to build the target URI on the fly from previous knowledge (user agents know a priori the WPS protocol to build valid URL request) and the pieces of information found in the capabilities response document (process identifier, service endpoint, version, etc.). User agents and services are coupled because they share the rules to build valid resource URI. This fact impedes that user agents and servers evolve independently since future changes in the service, either via an update in the specification or just a simple URI change, undoubtedly affect the implementation of client applications.

As the WPS service endpoint is only identified by an URI and none of the contained processes are exposed as resources, the first step towards RESTful geoprocessing services should be to identify each resource in a given WPS service. In doing so, user agents can automatically access to every single process resource in such a way that dereferencing its URL permit them to manipulate that resource.

Although some authors have proposed some rules for URI building (Mannens et al, 2012) (Janowicz et al., 2010), it is not expected that user agents have prior knowledge of these rules to access every resource. In contrast, a good REST design demands only a public URI, which acts as a normative or canonical URI, to let user agents to access any resource deployed in a server. Indeed, user agents should be able to discover related resources by querying this public URI, which becomes the single entry point to all geospatial resources and services within a GII node. By hiding the URI construction rules imposed uniquely by the server, user agents become more independent with respect to future changes in resource URI policies.

### 3.2.2 Representations

Resources are abstract entities that cannot be directly manipulated by user agents. Yet resources may be regarded as being a set of attributes and properties which are accessible and manipulable. A resource representation is then an informational view of a resource at a given instant in time (Fielding, 2000). Resource representations are a crucial aspect to embody the uniform interface concept tenet because user agents and services communicate each other by exchanging resource representations through the fixed set of HTTP methods (Section 3.1.4).

The separation between abstract entities and representations makes it possible a one-to-many relationship between a resource and their representations. Each representation is a different view of the same underlying resource. In addition, such representations are encoded at runtime into transferable formats through the HTTP's content negotiation mechanism (Fielding et al., 1999). The use of broadly accepted media types ensures that user agents are capable by default of understanding and manipulating resource representations.

Web services encourage the separation between service implementation and service interface (Papazoglou, 2008). This also occurs in the WPS–based services since all of the OGC specifications uniquely describe service interfaces. For this reason, user agents interact with remote geoprocessing services via messages (representations) during the request-response communication. A WPS service responds with a metadata document subject to the operation requested. So user agents get a representation, i.e., a projection of the current values of the operation requested.

The WPS specification defines XML schemas for responses and exceptions that are concerned with the results of the operation requested. Successful responses contain the output run either by value (embedded) or by reference. In the latter case, an HTTP URI points to the resulting data file. Responses contain embedded output data, links to remote data, or even, a single output value that may be returned without any XML message. This suggests that the WPS specification support a kind of content negotiation in terms of desirable format of the response, which user agents specify in the request.

Apart from the *execute* response, the remaining WPS operations do not support the ability to select the best representation of the result operation. For instance, a *describeProcess* response always comes in XML format, and user agents have no way to select an alternative representation format (e.g., JSON). Two user agents (user and a web map application) may consume any given WPS-based service, but both may have different needs in terms of representation formats. While a user would expect results in HTML format for reading, a web map application would expect other specific formats (JSON or KML) for automatic processing. Furthermore, the ability of demanding customized resource representations is not supported in WPS-based services.

### 3.2.3 Hypermedia

Terms such as application state and resource state are sometimes misunderstood (Foster et al., 2008). The space of all valid states at a given time that a user agent can choose from encompasses the application state. User agents need to know the valid interactions with a given set of resources to accomplish a certain business goal. As user agents make progress towards that goal, their application state changes and evolves accordingly. As a result, each user agent maintains its own application state that persists across several interactions (left side of Fig. 3).

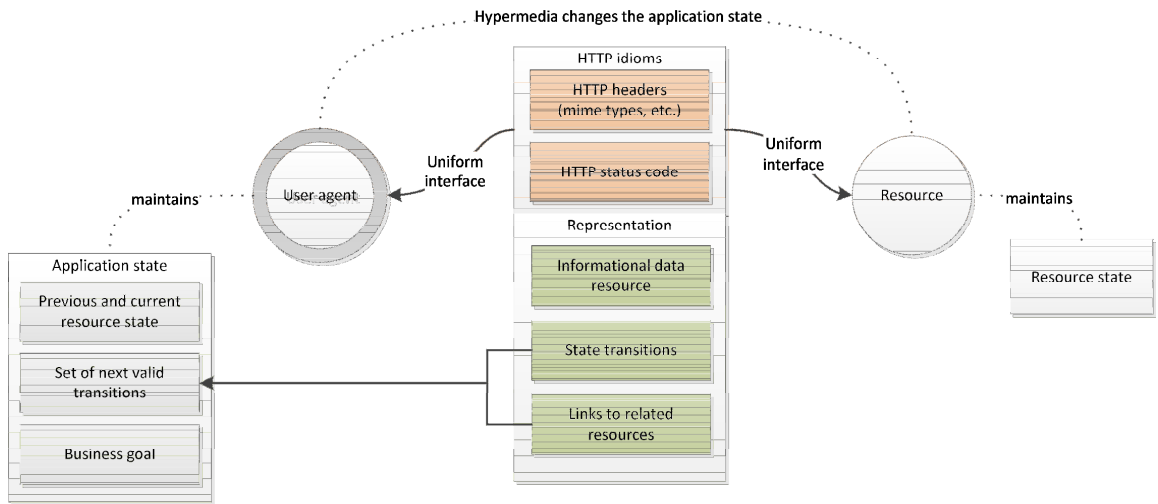

Figure 3. Hypermedia as a means to change the application state to pursue a goal

The term hypermedia refers to how the legal interactions between user agents and resources are built upon the combination of state transactions. Hypermedia is closely tied to the concept of link. When navigating, links within a given page connect related web pages and resources. As links are labelled, the understanding of these labels is shared between users and web page providers so that users may choose one of the set of links available (transitions) in the current web page (resource). In REST, state transitions are similar to the idea of labelled links on web pages. The list of state transitions available in a resource representation indicates the potential target resources to which a user agent may visit. Selecting the next transition produces changes in the application state maintained by the user agent (Figure 3).

Rather than using a simple textual description for each link as in the case of web browsing, user agents need accurate information about the legal state transitions to ease decision making. For this reason, state transitions are based on typed links and the meaning of each type is explicitly agreed and shared among the involved participants. Each link type (e.g., category) is in turn defined by the combination of accepted media types and supported HTTP methods and status codes. This information allows user agents to know in advance the media type, representation formats and exceptions of a given resource in order to properly interact with. In doing so, user agents maintain the application state correctly by discovering, making decisions, and following typed links at run time in a generic manner as we (humans) do when browsing.

In the WPS service domain, as input and outputs are well defined according to the WPS schema, process parameters may be readily encoded in a URI in the KVP approach. Input parameters of a given process are then constructed on-the-fly by calling to other WPS services or geospatial services (e.g., WFS or SOS as shown in Fig. 1). For instance, the input of the Area process could be a dereferenceable dataset (resource) as the result of intersecting two geometries. This approach facilitates service chaining because each service requested is uniquely identified via an HTTP URI. In contrast to XML/POST and XML/SOAP bindings, the KVP approach comes nearer to the REST style because the combination of a set of resources is much easier when every single resource is correctly identified (Section 3.2.1).

Service chaining may be seen as a strategy to explicitly describe a target goal, that is, the steps needed to accomplish that goal. Similarly, BPEL descriptions may be used also to put together processes from various WPS-based services (Chen et al., 2009). In any case, the chain of steps is imposed and known at design time. Opposite to service chaining at design time is the hypermedia. The hypermedia constraint promotes ad hoc, dynamic interactions discovered by the user agents as they interact with the resources. Nevertheless, user agents and services are tight coupled by contract imposed by the WPS specification protocol at design time. For instance user agents know that after receiving a *describeProcess* response it follows an *execute* call. From the REST perspective, user agents discover this kind of information by looking at one response at a time, to evaluate then how best to proceed given the available transitions.

As commented earlier, WPS responses contain XML messages with embedded output data or links to data files or raw values for single process output. This suggests that WPS responses support links to related output data yet, in any case, such links do not provide any information, in the sense of valid transitions, about how user agents may proceed after collecting the results. WPS responses do not include typed links to guide user agents whether this link should be chosen or not. For instance, let's suppose that the 'Area' process fails by a timeout exception. Currently, user agents are informed with an exception code and possibly a brief message that indicates the cause of the error. However, the lack of legal transitions in the response such as "simulate execution", "try similar processes", and "try again later" makes it difficult to figure out how to go on. Furthermore, current WPS clients are not able to know the next step to follow (e.g., invoking the 'Area' process, accessing data sets, etc.) by looking only at the response message of the Intersect process. User agents collect the resulting intersection geometry but pointers to related resources encoded in the very representation are missing.

Some works have recently addressed this issue to examine the challenges and opportunities in augmenting geospatial services descriptions with typed links (Schade et al., 2010; Lopez-Pellicer et al., 2010). They suggest the use of typed or semantic links to connect dereferenceable resources as promoted by Linked Data (Bizer et al., 2009). Allowing user agents to discover next state transitions through each request-response interaction promotes loosely-coupled applications because decisions on the application protocol are not imposed at design time but discovered at run time. In such a context, a simple improvement of the WPS specification could be to advertise typed links to the corresponding *describeProcess* operations within the *getCapabilities* response. The definition of such typed links is a challenging task (Brauner et al, 2009). For instance, Bai et al (2009) suggest the definition of an integrated taxonomy of geoprocessing functions to group and classify geospatial services. The use of taxonomic terms would also enable to establish connections between similar geoprocessing functions regardless of the name used to identify them.

Apart from links embedded in resource representations, media types are relevant to support hypermedia. The use of media types in WPS responses is exclusively considered to identify the format of a process input or output. For instance process outputs may come in GML or KML mime types, as defined in the *describeProcess* response. Mime types are thus applied neither to the whole representation nor to state transitions. Therefore, media types are not used to anticipate the format of the current resource or the expected format of a related resource pointed by a link. The correct use of mime types is an essential mechanism in REST to provide useful information about the state of a given resource.

Exceptions are also an important part of the application state (Foster et al., 2008). When an error occurs while invoking *describeProcess* or *execute* methods, the server should return an exception message indicating the cause of the error if possible. The WPS specification defines some error messages, which simplifies the task of error handling for multiple processes in the same WPS service. Although the idea of exception handling is present, the number of pre-defined exception types (e.g. *MissingParameterValue* and *ServerBusy)* seems quite limited. Michaelis and Ames (2009) also identified an inconsistency in the WPS specification after executing a process with regards to what a client should expect next. As a valid response might be either a XML response document or an Exception document, user agents are not able to know the status of the process run until the response message is parsed in the client side.

RESTful applications rely on the use of the rich semantics of the standard HTTP status codes. Indeed, the range of WPS exception codes has a possible correspondence with an HTTP status code. For instance, *MissingParameterValue* exception may be represented by 400 status code (bad request because of malformed URI) while 503 (request timeout) is suitable to indicate that the server is temporary busy to handle incoming requests. Building exception handling upon the standard HTTP status code mechanism aligns web applications to the Web architecture and takes for granted all its benefits.

### 3.2.4. Self-descriptive messages

In contrast to the application state seen previously, the resource state is exchanged in the interactions between user agents and resources through resource representations. Resource attributes, valid state transitions, and pointers to other related resources are often embedded in the resource representation are part of the resource state. In addition, HTTP headers also are part of the resource state because these metadata descriptors help user agents to interpret the meaning of the resource representation transferred in the HTTP body message. In this case, servers manage the state of their own resources (right side of Fig. 3).

Self-descriptive messages mean that the semantics of the exchanged messages is completely understandable and visible for any third party participant involved in the user agent-service communication. This is an immediate result of the application of the stateless constraint, since all that is needed to interpret the interactions between clients and services at run time is contained within those exchanged requests and response messages (Webber et al., 2010).

Although the separation between resource and representation is somehow encouraged (see Section 3.2.1), the content of the WPS responses contain nothing else than the attributes of the process. A representation resource should provide a comprehensive view of the client-server communication at a given instant of time, rather than being constrained exclusively to the results of an operation run. For instance, apart from getting a description of the Area process, user agents may be interested in knowing input datasets already uploaded in the server to use directly in the process. This kind of information could be included in the WPS response. Furthermore, the role of representations in current WPS services is basically aimed to give informational resource data rather than providing additional bits of contextual information such as state transitions and pointers to related resources that would help user agents in interpreting the current resource representation.

Self-descriptive representations provide several benefits. Firstly, legacy applications can be encapsulated behind services as current WPS-based services do. Secondly, mediators may understand messages between clients and servers because their meaning is explicit to all. In contrast to WPS services, specialized mediators such as coordinate transformations, filtering or caching could take part in the request-response communication. Thirdly, the real benefit from the WPS services' perspective is that client and server implementations may evolve at different rhythms without depending each other. As each type of participant (user agents, services, and mediators) is able to interpret the semantics of the resource representations at run time, updates and changes over time are explicitly and immediately understood. Self-descriptive representations together with stateless communication patterns minimize coupling and maximize scalability (Fielding, 2000).

# 4. Discussion

The goal of the paper was to determine whether current WPS-based geoprocessing services can be viewed under the lens of the REST architectural style. To meet this objective we have analyzed the WPS specification from the set of REST constraints, i.e., client-server, stateless, cache, uniform interface, layered approach, and code-on-demand, and the REST data elements, i.e., identification of resources, representation, hypermedia and self-descriptive messages, which are behind some REST constraints.

A summary of the discussion conducted in the paper is shown in Tables 1 and 2. The assessment results suggest that WPS geoprocessing services do not resemble RESTful services in every constraint. The major differences are the identification of resources and the use of HTTP as a transport protocol to run RPC-based methods. Future developments towards RESTful geoprocessing services should align at least to the uniform interface constraint (and the data elements behind this constraint) to actually consider HTTP as an application protocol. Indeed, many characteristics of the HTTP protocol such as media types, content-negotiation, status codes and HTTP headers, may be applied directly to WPS-based services. This would avoid the burden of similar capabilities of the WPS specification that HTTP protocol already supports.

We believe that exploiting the full potential of HTTP in geospatial processing applications would greatly simplify not only the WPS specification but OGC specifications in general because common functionalities will be based entirely on built-in HTTP mechanisms. The combination of the REST constraints and data elements described here plays an important role in establishing effective connections among resources (Erenkrantz et al., 2007; Krummenacher et al., 2010). Research efforts towards the creation of RESTful services on top of current GII resources should be encouraged to increment not only the amount of internal links among GII resources but to enhance the connections with resources in other domains (Granell et al., 2011). Recent research works are addressing this issue (Gao et al., 2010; Mazzetti et al., 2010), although the experiments conducted do not cope with all the REST constraints described here. Whilst the identification of resources and the use of HTTP methods are normally well understood, the application of typed links, content-negotiation and hypermedia controls is not regarded yet.

The use of the REST approach to model geoprocessing services does not imply to oversimplify the complexity and variability present in the geospatial service domain (Tamayo et al., 2012), but to provide a complementary vision to deal with it. Some approaches may be envisioned for designing RESTful geoprocessing services. One is to update the current WPS specification to support the set of REST constraints by adding a 'REST binding'. This may be inappropriate because the underlying service model stands invariable and still mirrors to the application-specific WPS interfaces. Changes in the underlying data model are necessary to move from a service-oriented model to a resource-oriented model (Pautasso et al., 2008).

Another proposal would be (currently we are exploring this approach) to maintain intact existing geoprocessing services, mainly in terms of OGC WPS specifications, and adding a light RESTful

API that act as a mediator for such WPS geoprocessing services within GII nodes. In doing so, the resulting RESTful geoprocessing services would gain in connectiveness since such process resources would provide native support for typed link relations and be flexible to accommodate new paths between GII nodes that have never existed before. Also, composing various geoprocessing services becomes a matter of following links in order for user agents to discover and follow valid state transitions at retrieval of resource representations. In web integration scenarios where a great variety of user profiles interact with geospatial services, ad hoc composition of RESTful services becomes a need.

## 5. Conclusion

This research makes an attempt to provide insights into the mechanisms and functionality of WPS-based geoprocessing services from the perspective of REST. Although current service communication and binding styles do not fully suit in genuine RESTful solutions, most of the concepts and characteristics could be converted into a resource-oriented approach, which may be the way to enhance scalability, statelessness and hypermedia into geoprocessing services.

The value of decoupling resources from representations, the assimilation of HTTP application protocol and the use of typed link relations are of crucial importance to accommodate *ad hoc* composition of geoprocessing services. Future advances in geospatial workflow modelling and data mining scenarios would benefit from adopting a RESTful approach.

| REST constraint | KVP Protocol binding | XML/POST protocol binding | XML/SOAP protocol binding | Desirable (RESTful) requirements |
|---|---|---|---|---|
| Client-Server | **Yes**. Decoupling client and service roles is a crucial aspect in Web services in general and, hence, in OGC services. | | | Separation of concerns is encouraged by client-server constraint. |
| Stateless | **Partially Yes**. A given interaction does not depend on previous or future interactions. However, self-descriptiveness does not stem from the stateless communication in the realm of WPS services. | | | No shared session. Application state managed by user agents. |
| Cache | **No**. The use of HTTP headers would enable cache, which requires considering HTTP as application protocol. | | | HTTP's cache mechanism used correctly. |
| Uniform interface | **No**. RPC style and HTTP as transport protocol invalid this tenet. Variable number of operations because such operations are imposed by the WPS specification. | | | Semantics of HTTP verbs is correctly used to manipulate resources. Fix operation set whatever the application domain. |
| | **No**. Operation semantics depend on the method tunneled through the *request* parameter within URI. | **No**. Operation semantics depend on the method contained in the body payload as an XML document. | **No**. Operation semantics depend on the method contained in the SOAP body as an XML document. | Semantics of HTTP verbs used correctly to manipulate exposed resources. |
| Layered approach | **Partially Yes**. WPS services play the role of intermediaries on layered architectures, but lack the ability to transform and manage self-descriptive messages. | | | Self-descriptive resource representations promote the use of intermediaries. |
| Code-on-demand | **No**. Algorithms are always executed at servicer side. | | | Downloading code to clients is encouraged. |

Table 1. Summary assessment of the support of REST constraints within WPS services

| REST Data elements | Aspects | KVP Protocol binding | XML/POST protocol binding | XML/SOAP protocol binding | Desirable (RESTful) requirements |
|---|---|---|---|---|---|
| **Self-descriptive messages** | Resource data | **No**. Representations only contain an information view of the operation requested. For instance, hypermedia controls are missing. | | | Resource representations contain all needed information (data, typed links, pointers to related resources) to be understood by user agents, services and intermediaries. |
| | State transitions | **No**. State transitions (typed links) are not included in resource representations. Assumptions about valid transitions are known at design time. | | | |
| | Pointers to related resources | **Partially yes.** Responses may contain a pointer to the output results. | | | |
| **Identification of resources** | Abstraction identification | Resource | Service | Service | Resource |
| | | **Yes**. Resources are identified through URI. | **No**. Only the service endpoint is identified by URI. Other resources are hidden. | **No**. Only the service endpoint is identified by URI. Other resources are hidden. | Resources are exposed and identified correctly using HTTP's URI mechanism. |
| | Canonical URI | **No**. No entry-point URI to access related resources. | **Partially yes.** Service endpoint URI may be used as canonical URI. | **Partially yes.** Service endpoint URI may be used as canonical URI. | Canonical URI as entry point is encouraged. |
| **Representation** | Separation | **Yes.** Real entities and informational representations are loose coupled, as web services define. | | | Access to resources is mediated by their representations. |
| | Content-negotiation | **Partially yes**. Only *execute* responses are subject to some sort of content negotiation with WPS own mechanisms. HTTP content negotiation is not used. | | | HTTP content negotiation is encouraged to select the best format representations on the same resource depending on client's needs. |
| **Hypermedia** | Dynamic interactions | **No**. URI space and interactions known at design time. | **No**. Interactions known at design time. | **No**. Interactions known at design time. | User agents look at one response at a time, to evaluate how to proceed given the available transitions. |
| | Use of typed links | **No**. Valid transitions are known at design time. | | | Next valid transitions are selected at runtime based on typed links and current application state. |
| | Use of media types | **No**. Available representation format is known at design time. Only media types for output parameters may be selected. | | | Generic and specific media types used correctly. Resource representations and transitions based on |

| | | | |
|---|---|---|---|
| | | | media types. |
| | External metadata | **No**. No additional metadata about the message as a whole. | HTTP headers codify extra pieces of information about the representation of the resource. |
| | Support of exception handling | **Partially yes**. A defined XML interface for exceptions. However, it does not rely .on HTTP mechanisms. | HTTP status codes define a great range of success and failure conditions. |
| | Use of HTTP idioms | **No**. HTTP used as transport protocol. | HTTP headers, HTTP status codes, and HTTP verbs form part of the application protocol. |

Table 2. Summary assessment of the support of REST data elements within WPS services